\documentclass{PoS}

\usepackage{amsfonts, amsmath, amssymb}
\usepackage{graphicx}
\usepackage{booktabs, multirow}
\usepackage{siunitx}
\usepackage{hyperxmp}
\usepackage{color}
\usepackage{simplewick}

\usepackage{relsize}
\newcommand{\babar}{\mbox{\slshape B\kern-0.1em{\smaller A}\kern-0.1em B\kern-0.1em{\smaller A\kern-0.2em R}}}

\newcommand{\VMD}{\mathrm{VMD}}
\newcommand{\LMD}{\mathrm{LMD}}
\newcommand{\LMDV}{\mathrm{LMD+V}}
\newcommand{\dof}{\mathrm{d.o.f.}}

\newcommand{\fm}{\mathrm{fm}}
\newcommand{\MeV}{\mathrm{MeV}}
\newcommand{\GeV}{\mathrm{GeV}}

\newcommand{\psib}{\overline{\psi}}

\newcommand{\FF}{{\cal F}_{\pi^0\gamma^*\gamma^*}}

\newcommand{\amu}{a_\mu^{\mathrm{HLbL}; \pi^0}}

\title{Lattice calculation of the pion transition form factor $\pi^0 \to \gamma^* \gamma^*$}

\ShortTitle{Lattice calculation of the pion transition form factor $\pi^0 \to \gamma^* \gamma^*$}

\author{\speaker{Antoine G\'erardin}\\
        PRISMA Cluster of Excellence and Institut f\"ur Kernphysik, Johannes Gutenberg-Universit\"at Mainz, 55099 Mainz, Germany\\
        E-mail: \email{gerardin@kph.uni-mainz.de}}

\author{Harvey B. Meyer\\
        PRISMA Cluster of Excellence and Institut f\"ur Kernphysik, Johannes Gutenberg-Universit\"at Mainz, 55099 Mainz, Germany\\
        Helmholtz Institute Mainz, Johannes Gutenberg-Universit\"at Mainz, 55099 Mainz, Germany\\
        E-mail: \email{meyerh@uni-mainz.de}}

\author{Andreas Nyf\/feler\\
        PRISMA Cluster of Excellence and Institut f\"ur Kernphysik, Johannes Gutenberg-Universit\"at Mainz, 55099 Mainz, Germany\\
         E-mail: \email{nyffeler@kph.uni-mainz.de}}
        
\abstract{We calculate the pion transition form factor $\FF(q_1^2,q_2^2)$, which describe the interaction of an on-shell pion with two off-shell photons, using lattice QCD simulations with two degenerate flavors of dynamical quarks. This form factor is the main ingredient in the calculation of the pion-pole contribution to hadronic light-by-light scattering in the muon $g-2$, $\amu$. We focus our study on the spacelike region with photon virtualities up to $1.5~\GeV^2$, not yet measured experimentally. Several lattice spacings and pion masses are used to extrapolate the results to the physical point and a comparison with different phenomenological models is performed. Finally, we use our extrapolated form factor to provide a lattice determination of $\amu$.}

\FullConference{34th annual International Symposium on Lattice Field Theory\\
		24-30 July 2016\\
		University of Southampton, UK}

\begin{document}

\section{Introduction} 
\label{sec:intro}

The anomalous magnetic moment of the muon provides one of the most precise tests of the Standard Model of particle physics~\cite{Jegerlehner:2009ry,Miller:2012opa} and a persistent discrepancy of about $3-4$ standard deviations~\cite{Agashe:2014kda} exists between experiment and theory. In the near future, the experimental error is expected to be reduced by a factor four~\cite{future_g-2_exp}. The theoretical error is now dominated by hadronic contributions : the hadronic vacuum polarization (HVP) and hadronic light-by-light scattering (HLbL) and, for the latter, no reliable estimate exists yet and systematic errors are difficult to estimate.
However, recently a dispersive approach was proposed~\cite{HLbL_DR} which relates the numerically dominant pseudoscalar-pole contribution, and the pion-loop in HLbL with on-shell intermediate pseudoscalar states to measurable form factors and cross-sections with off-shell photons: $\gamma^* \gamma^* \to \pi^0, \eta, \eta^\prime$ and $\gamma^* \gamma^* \to \pi^+ \pi^-, \pi^0 \pi^0$. Within this framework, the pion-pole contribution is obtained by integrating some weight functions times the product of a single-virtual and a double-virtual transition form factors for spacelike momenta~\cite{Jegerlehner:2009ry}. In particular, the weight functions turn out to be peaked at low momenta such that the main contribution to $a_{\mu}^{{\rm HLbL}; \pi^0}$ arises from photon virtualities below $1~\GeV^2$ \cite{KN_02_Nyffeler:2016gnb}, a kinematical range accessible on the lattice.
From the experimental point of view, only the single-virtual form factor for the pion ${\cal  F}_{\pi^0\gamma^*\gamma^*}(-Q^2,0)$  has been measured \cite{exp} in the spacelike region $Q^2 \in [0.5, 40]~\GeV^2$. From the theoretical point of view, the form factor is constrained by the Adler-Bell-Jackiw (ABJ) anomaly in the chiral limit such that $\FF(0, 0) = 1/(4\pi^2 F_{\pi})$ \cite{adler_bell_anomaly}. The single-virtual form factor has been computed in the framework of factorization in QCD (operator-product expansion (OPE) on the light-cone) and one finds the Brodsky-Lepage behavior~\cite{BL_3_papers} 
\begin{equation}  
\FF(-Q^2, 0) \xrightarrow[Q^2 \to \infty]{}   2 F_{\pi}  / Q^2 \,.  
\label{eq:BL}
\end{equation} 
Finally, the double-virtual form factor where both momenta become
simultaneously large has been computed using the OPE at short
distances. In the chiral limit the result
reads~\cite{Nesterenko:1982dn, Novikov:1983jt}
\begin{equation} 
\vspace{-0.2cm}
\FF(-Q^2, -Q^2) \xrightarrow[Q^2 \to \infty]{}  2 F_{\pi} / (3 Q^2) \,. 
\label{eq:OPE}
\end{equation} 
Therefore, the double-virtual form factor in the kinematical range of interest $[0-1]~\GeV^2$ for
the computation of the HLbL contribution to the muon $g-2$ is still unknown and the available estimates rely on phenomenological models \cite{Jegerlehner:2009ry, Bijnens:2015jqa}. Previous lattice studies \cite{Lin:2013im_Feng:2012ck} < on the decay $\pi^0 \to \gamma\gamma$ (form factor at very low momenta). More details on this work can be found in \cite{Gerardin:2016cqj}.

\section{Methodology \label{sec:methodology}}

In Minkowski spacetime, the form factor of interest is defined via the following matrix element
\begin{equation} 
\vspace{-0.2cm}
M_{\mu\nu}(p,q_1)  = i \int \mathrm{d}^4 x \, e^{i q_1 x} \, \langle \Omega | T \{ J_{\mu}(x) J_{\nu}(0) \} | \pi^0(p) \rangle =
\epsilon_{\mu\nu\alpha\beta} \, q_1^{\alpha} \, q_2^{\beta} \, \FF(q_1^2, q_2^2) \,, 
\label{eq:MFF}
\end{equation}
where $q_1$ and $q_2$ are the photon momenta and $p = q_1 + q_2$ is the on-shell pion momentum. $J_{\mu} = \sum_f Q_f \, \psib_f \gamma_{\mu} \psi_f$ is the hadronic component of the electromagnetic current and we use the relativistic normalization of states $\langle \pi^0(p) | \pi^0(p^{\prime}) \rangle = (2\pi)^3\, 2 E_{\pi}(\vec{p}) \ \delta^{(3)}(\vec{p}-\vec{p}^{\  \prime})$. To compute the form factor on the lattice, we follow the method introduced in \cite{Ji:2papers_dudek}. Keeping $q_{1,2}^2 < M^2_V = {\rm min}(M^2_{\rho}, 4m_\pi^2)$, one can show \cite{Feng:2012ck} that the matrix element in Euclidean spacetime is
\begin{equation}
M_{\mu\nu}  =  (i^{n_0}) M_{\mu\nu}^{\rm E}, \quad 
M_{\mu\nu}^{\rm E}  \equiv  - \int \mathrm{d} \tau \, e^{\omega_1 \tau}  \int \mathrm{d}^3 z \, e^{-i \vec{q}_1 \vec{z}} \, 
\langle 0 | T \left\{ J_{\mu}(\vec{z}, \tau) J_{\nu}(\vec{0}, 0) \right\} | \pi(p) \rangle  \,,
\label{eq:Mminkowski}
\end{equation}
where $\omega_1$ is a real free parameter such that $q_1 = (\omega_1, \vec{q}_1)$ and $n_0$ denotes the number of temporal indices carried by the two vector currents.  Therefore, one is led to consider the following three-point correlation function on the lattice
\begin{align}
C^{(3)}_{\mu\nu}(\tau,t_{\pi}) = a^6\sum_{\vec{x}, \vec{z}} \, \big\langle  T \left\{ J_{\mu}(\vec{z}, t_i) J_{\nu}(\vec{0}, t_f)  P^{\dag}(\vec{x},t_0) \right\} \big\rangle \, e^{i \vec{p}\, \vec{x}} \, e^{-i \vec{q}_1 \vec{z}} \,,
\label{eq:lat_cor}
\end{align}
where $\tau=t_i-t_f$ is the time separation between the two vector currents and  $t_{\pi}={\rm min}(t_f-t_0,t_i-t_0)$. The matrix element with on-shell pion is obtained by considering the large $t_{\pi}$ limit. By defining
\begin{equation}
A_{\mu\nu}(\tau) = \lim_{t_{\pi} \rightarrow +\infty} C^{(3)}_{\mu\nu}(\tau,t_{\pi}) \, e^{E_{\pi}t_{\pi}}  \,, \quad 
\widetilde{A}_{\mu\nu}(\tau) = \left\{\begin{array}{l@{~~~}l}  A_{\mu\nu}(\tau) & \tau >0 \\  A_{\mu\nu}(\tau) \,  e^{-E_{\pi} \tau} & \tau<0 \end{array} \right.\;,
\label{eq:Amunu}
\end{equation}
and using Eq.~(\ref{eq:Mminkowski}), $M_{\mu\nu}$ can be obtained via
\begin{gather}
 M_{\mu\nu}^{\rm E} =  \frac{2 E_{\pi}}{ Z_{\pi} }  \left( \int_{-\infty}^{0} \, \mathrm{d}\tau \, e^{\omega_1 \tau} \, A_{\mu\nu}(\tau) \,  e^{-E_{\pi} \tau} +  \int_{0}^{\infty} \, \mathrm{d}\tau  \, e^{\omega_1 \tau} \, A_{\mu\nu}(\tau)  \right)  
=   \frac{2 E_{\pi}}{ Z_{\pi} }  \int_{-\infty}^{\infty} \, \mathrm{d}\tau \, e^{\omega_1 \tau} \, \widetilde{A}_{\mu\nu}(\tau) \,,
\label{eq:lat_M}
\end{gather}
where the overlap factor $Z_{\pi}$ and the pion energy can be extracted from the asymptotic behavior of the two-point pseudoscalar correlation function.

\section{Lattice computation \label{sec:lattice_simulation}}

\begin{figure}[t]

	\vspace{-0.5cm}

	\begin{minipage}[c]{0.49\linewidth}
	\centering 
	\includegraphics*[width=0.90\linewidth]{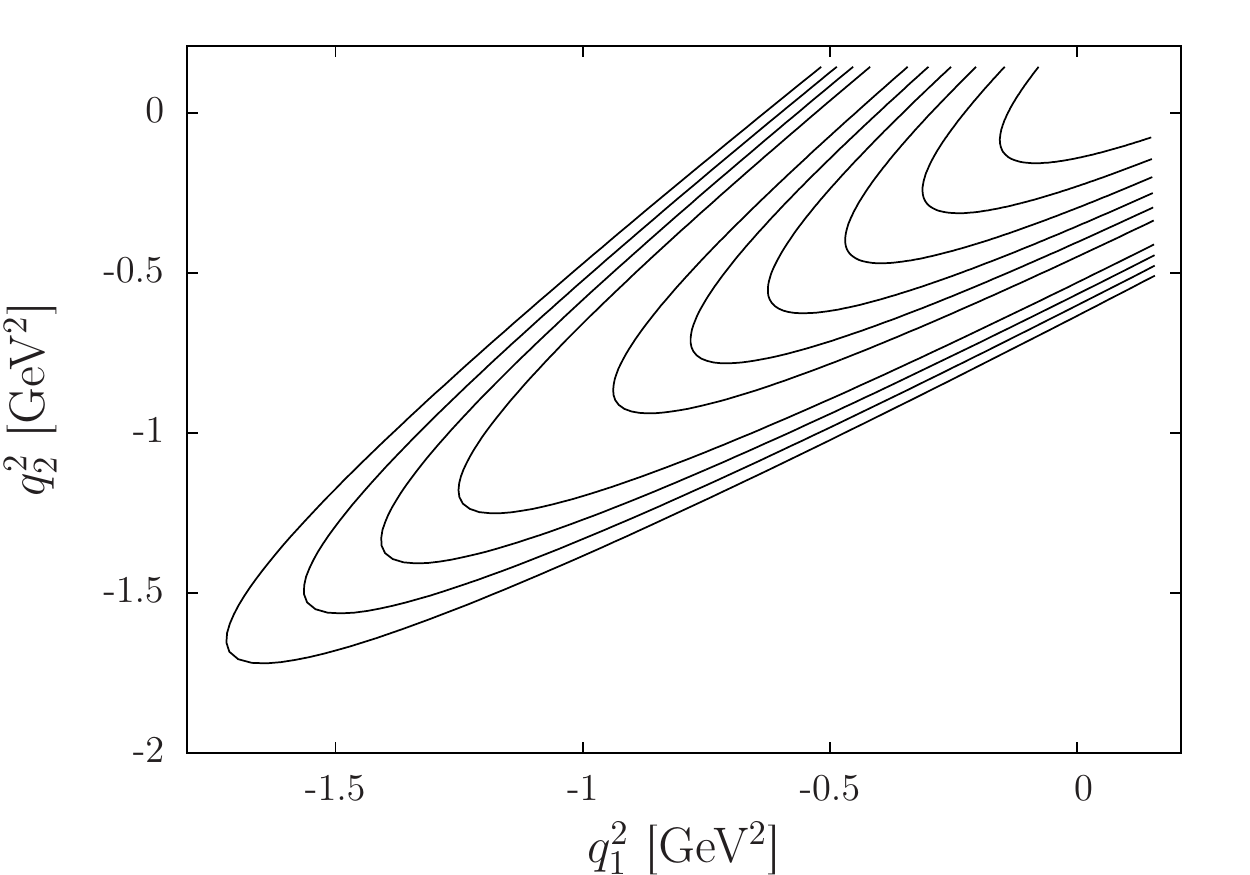}
	\end{minipage}
	\begin{minipage}[c]{0.49\linewidth}
	\centering 
	\includegraphics*[width=0.90\linewidth]{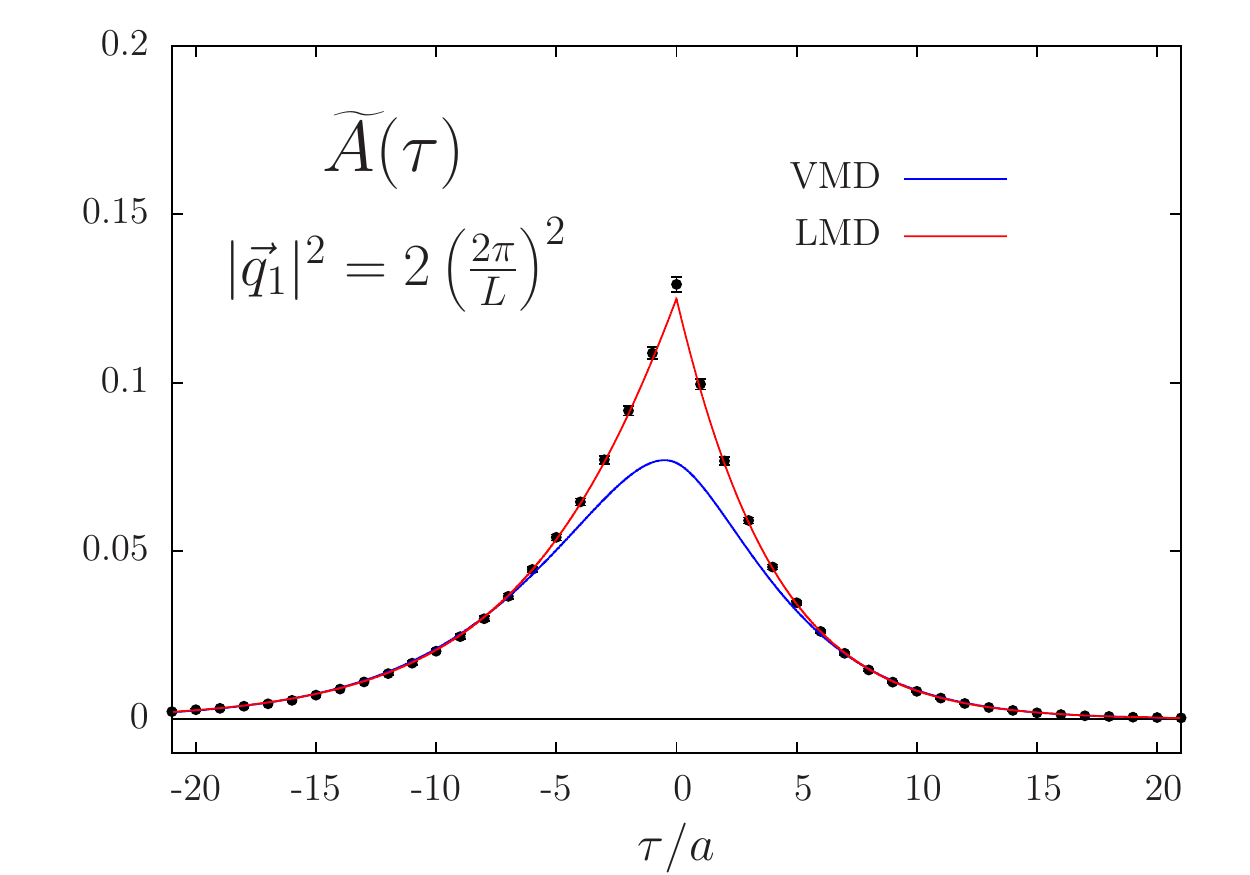}
	\end{minipage}
	\vspace{-0.2cm}
	\caption{(left) Kinematic reach in the photon virtualities ($q_1^2,q_2^2$) in our setup with the pion at rest, for the lattice resolution $48^3\times96$ at $a=0.065\,{\rm fm}$. (Right) The function $\widetilde{A}(\tau)$ (black points) and the VMD (blue line) and LMD (red line) fits used to describe the tail of the function at large $\tau$ for the lattice ensemble F7.}	
	\vspace{-0.2cm}
	\label{fig:kin}
\end{figure}

This work is based on a subset of the $n_f=2$ CLS (Coordinated Lattice Simulations) ensembles generated using the nonperturbatively $\mathcal{O}(a)$-improved Wilson-Clover action for fermions and the plaquette gauge action for  gluons. As shown in Table~\ref{tabsim}, three lattice spacings in the range [0.05-0.075]~fm are considered with pion masses down to 193~MeV and $Lm_{\pi}>4$ such that volume effects are expected to be negligible \cite{Meyer:2013dxa}. For more details on the ensembles, see~\cite{Fritzsch:2012wq}. The connected part of the three-point correlation function in Eq.~(\ref{eq:lat_cor}) has been computing using one `local' vector current $J_{\mu}^{l}(x) = \sum_f Q_f \ \overline{\psi}_f(x) \gamma_{\mu} \psi_f(x)$ and one `point-split' vector current 
\begin{align}
J_{\mu}^{c}(x) = \sum_f \frac{Q_f}{2} \left( \overline{\psi}_f(x+a\hat{\mu})(1+\gamma_{\mu}) U^{\dag}_{\mu}(x) \psi_f(x) - \overline{\psi}_f(x) (1-\gamma_{\mu} ) U_{\mu}(x) \psi_f(x+a\hat{\mu}) \right) \,,
\end{align}  
whereas the disconnected part is computed using two local vector currents. In the $\mathcal{O}(a)$-improved theory, the renormalized currents read $J_{\mu}^{\alpha,R}(x) = Z^{\alpha}_V (1+b_V^\alpha(g_0) am_q) \left(J_{\mu}^{\alpha}(x) + ac_V^\alpha \partial_{\nu} T_{\mu\nu} \right)$ with $\alpha=(\rm{local}, \rm{conserved})$ and where $b_V^\alpha$ and $c_V^\alpha$ are improvement coefficients. The point-split vector current satisfies the Ward identity and does not need any renormalization factor: $Z_V^{c,I} = 1$, $b_V^{c,I} = 0$ whereas $Z_V^{l}$ has been computed non-perturbatively in \cite{DellaMorteRD,Fritzsch:2012wq}. We neglect the contribution from the tensor density $T_{\mu\nu}(x)$ such that $\mathcal{O}(a)$-improvement is only partially implemented.  We choose the pion reference frame, $\vec{p}=0$, where both photons have back-to-back spatial momenta ($\vec{q}_2 = - \vec{q}_1$) and the kinematical range accessible on the lattice can be parametrized by
\begin{align}
\nonumber q_1^2 = \omega_1^2 - \vec{q}_1^{\, 2} \,, \quad q_2^2 = (m_{\pi} - \omega_1)^2 - \vec{q}_1^{\,2} \,.
\label{eq:kin}
\end{align}
We consider multiple values of $\vec{q}_1$ to obtain virtualities up to $|q_{1,2}^2| \approx 1.5~\GeV^2$ as can be seen in Fig.~\ref{fig:kin}. In this kinematical setup and using the Lorentz structure of the form factor one can show that only the spatial components are non-zero and can be written
\begin{equation}
\label{eq:Ascalar}
A_{kl}(\tau) = -i q_{kl} \, A(\tau)\,, \qquad q_{kl}\equiv \epsilon_{kl\alpha\beta} \, q_1^{\alpha} \, q_2^{\beta} = m_\pi\,\epsilon_{kli}\,q_1^i\,,
\end{equation}
where $A(\tau)$ is a scalar under the spatial rotation group ($\widetilde A(\tau)$ is defined in the same way). 

\begin{table}[t]
\caption{Parameters of the simulations: the bare coupling $\beta = 6/g_0^2$, the lattice resolution, the hopping parameter $\kappa$, the lattice spacing $a$ in physical units extracted from \cite{Fritzsch:2012wq}.}
\vskip 0.1in
\renewcommand{\arraystretch}{0.9}
\begin{tabular}{lcl@{\hskip 02em}l@{\hskip 01em}ccccccc}
	\hline
	CLS	&	$\quad\beta\quad$	&	$L^3\times T$ 		&	$\kappa$		&	$a~(\fm)$	&	$m_{\pi}~(\MeV)$	&	$F_{\pi}~(\MeV)$	& $m_{\pi}L$	&	$\#$confs \\
	\hline
	A5	&		$5.2$		&	$32^3\times64$	& 	$0.13594$	& 	$0.0749(8)$  	& 	$334(4)$ &	$106.0(6)$ &4.0&400 \\  
	B6	&					&	$48^3\times96$	& 	$0.13597$	& 		  		& 	$281(3)$ &	$102.3(5)$ &5.2& 400\\  
	\hline
	E5	&		$5.3$		&	$32^3\times64$	& 	$0.13625$	& 	$0.0652(6)$  	& 	$437(4)$ &	$115.2(6)$ &4.7 &400\\  
	F6	&		 	 		& 	$48^3\times96$	&	$0.13635$	& 				& 	$314(3)$ &	$105.3(6)$ &5.0&300 \\      
	F7	&		 	 		& 	$48^3\times96$	&	$0.13638$	& 				& 	$270(3)$ &	$100.9(4)$ &4.3&350 \\      
	G8	&		 	 		& 	$64^3\times128$	&	$0.136417$	& 				& 	$194(2)$ &	$95.8(4)$  &4.1&300\\    
	\hline  
	N6	&		$5.5$ 		& 	$48^3\times96$	&	$0.13667$	& 	$0.0483(4)$	& 	$342(3)$ &	$105.8(5)$ &4.0 &450\\      
	O7	&					& 	$64^3\times128$	&	$0.13671$	& 				& 	$268(3)$ &	$101.2(4)$ &4.2 & 150 \\      
	\hline
 \end{tabular} 
\label{tabsim}
\end{table}

\section{Results}

\subsection{Extraction of the form factor}

In Eq.~(\ref{eq:lat_M}), the time integration is performed using numerical data up to $\tau_c \approx 1.3~\fm$. For $\tau > \tau_c$, the contribution of the tail is estimated from a fit of our data with the analytical expression of $A^{\VMD}_{kl}(\tau)$ in the vector meson dominance model (VMD), derived in \cite{Gerardin:2016cqj} (see the next subsection for a description of the models). A typical fit for the lattice ensemble F7 is depicted in the right panel of Fig.~\ref{fig:kin} where the result using the lowest meson dominance model (LMD)~\cite{Moussallam:1994xp} rather that the VMD is also shown. Finally, the disconnected contribution to the three-point correlation function has been computed for the lattice ensemble E5 and only for the first three values of the spatial momentum $|\vec{q}_1|^2 = n^2 (2\pi/L)^2$, $n^2=1,2,3$. It contributes to less than $1\%$ of the total contribution and we conclude that the disconnected contribution is negligible at our level of accuracy.

\subsection{Fits in four-momentum space \label{subsec:fit_res}}

\begin{figure}[t!]
	
	\vspace{-0.5cm}

	\begin{minipage}[c]{0.32\linewidth}
	\centering 
	\includegraphics*[width=0.95\linewidth]{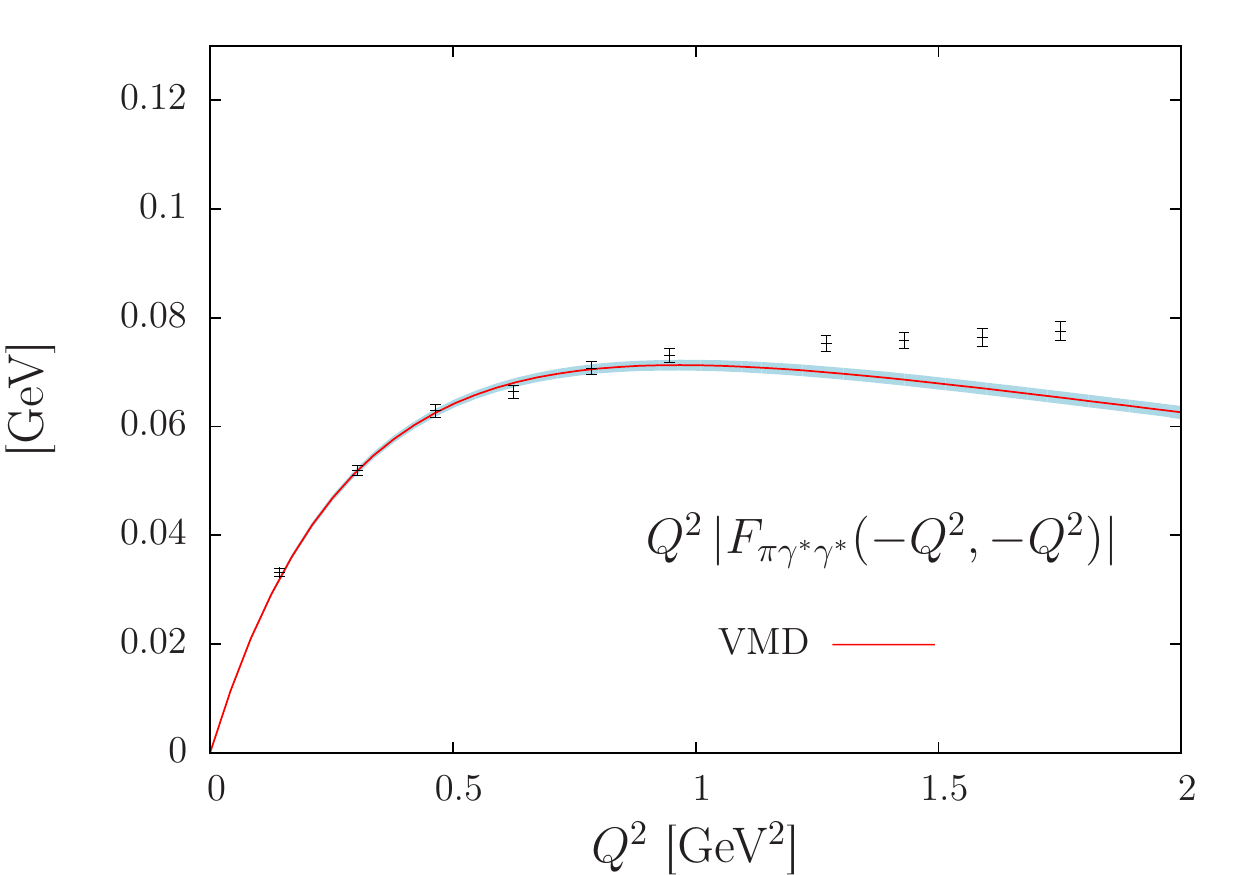}
	\end{minipage}
	\begin{minipage}[c]{0.32\linewidth}
	\centering 
	\includegraphics*[width=0.95\linewidth]{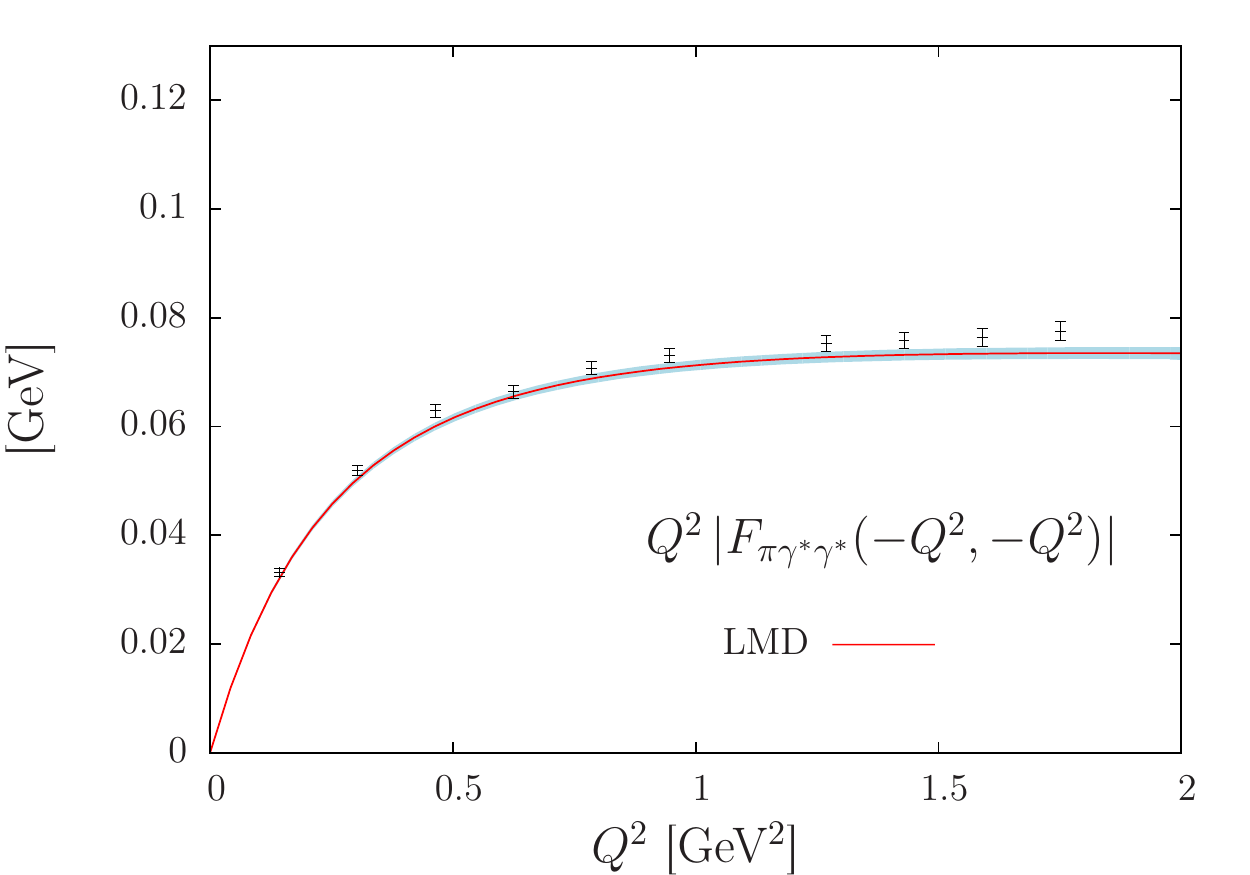}
	\end{minipage}
	\begin{minipage}[c]{0.32\linewidth}
	\centering 
	\includegraphics*[width=0.95\linewidth]{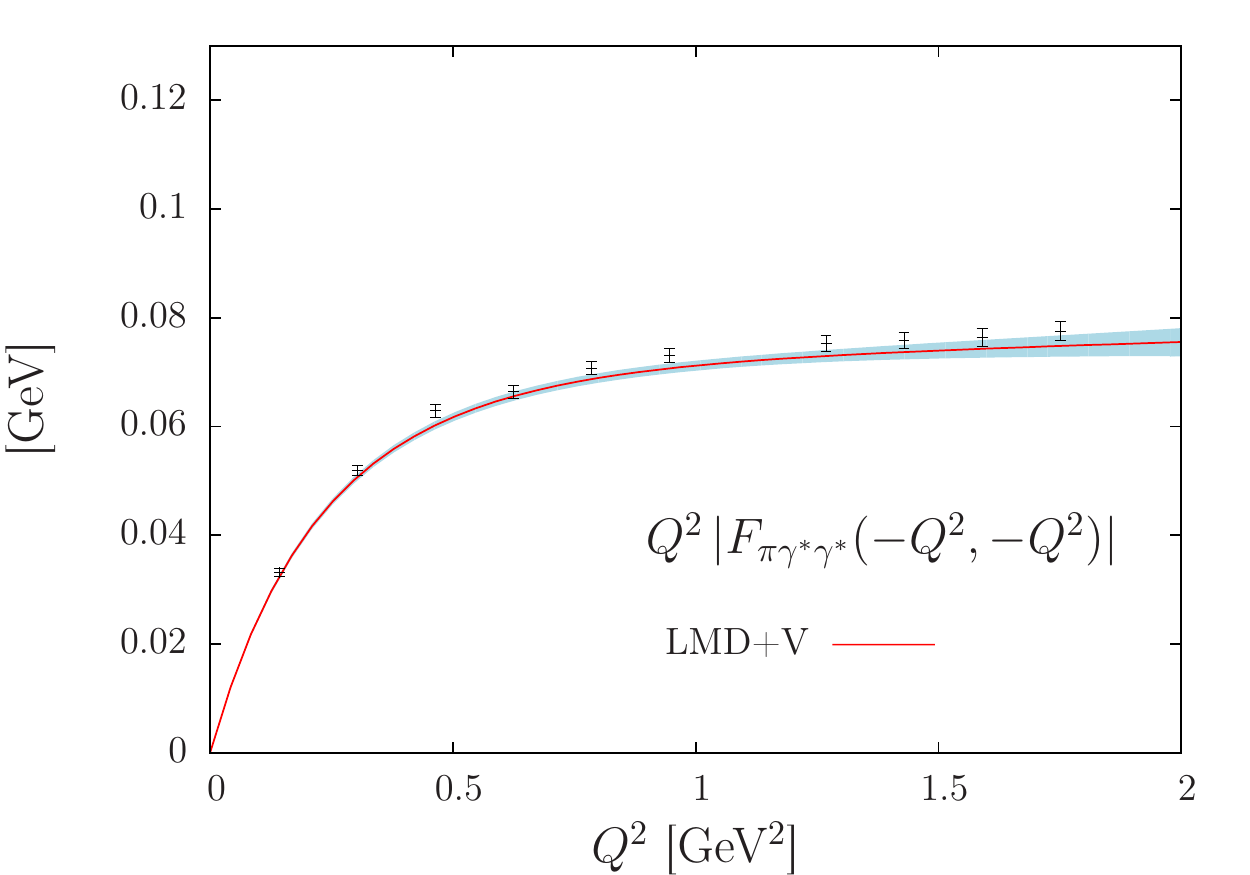}
	\end{minipage}
	
	\begin{minipage}[c]{0.32\linewidth}
	\centering 
	\includegraphics*[width=0.95\linewidth]{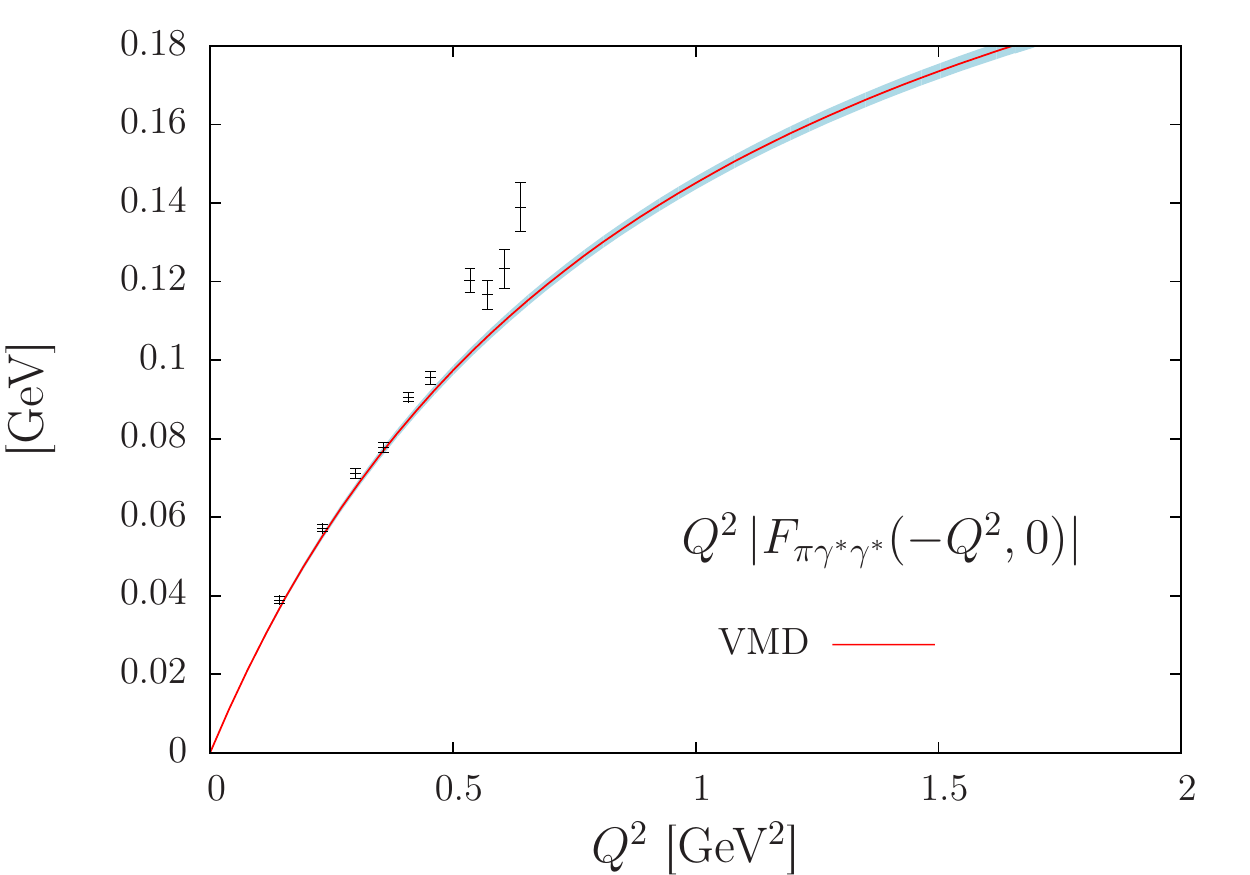}
	\end{minipage}
	\begin{minipage}[c]{0.32\linewidth}
	\centering 
	\includegraphics*[width=0.95\linewidth]{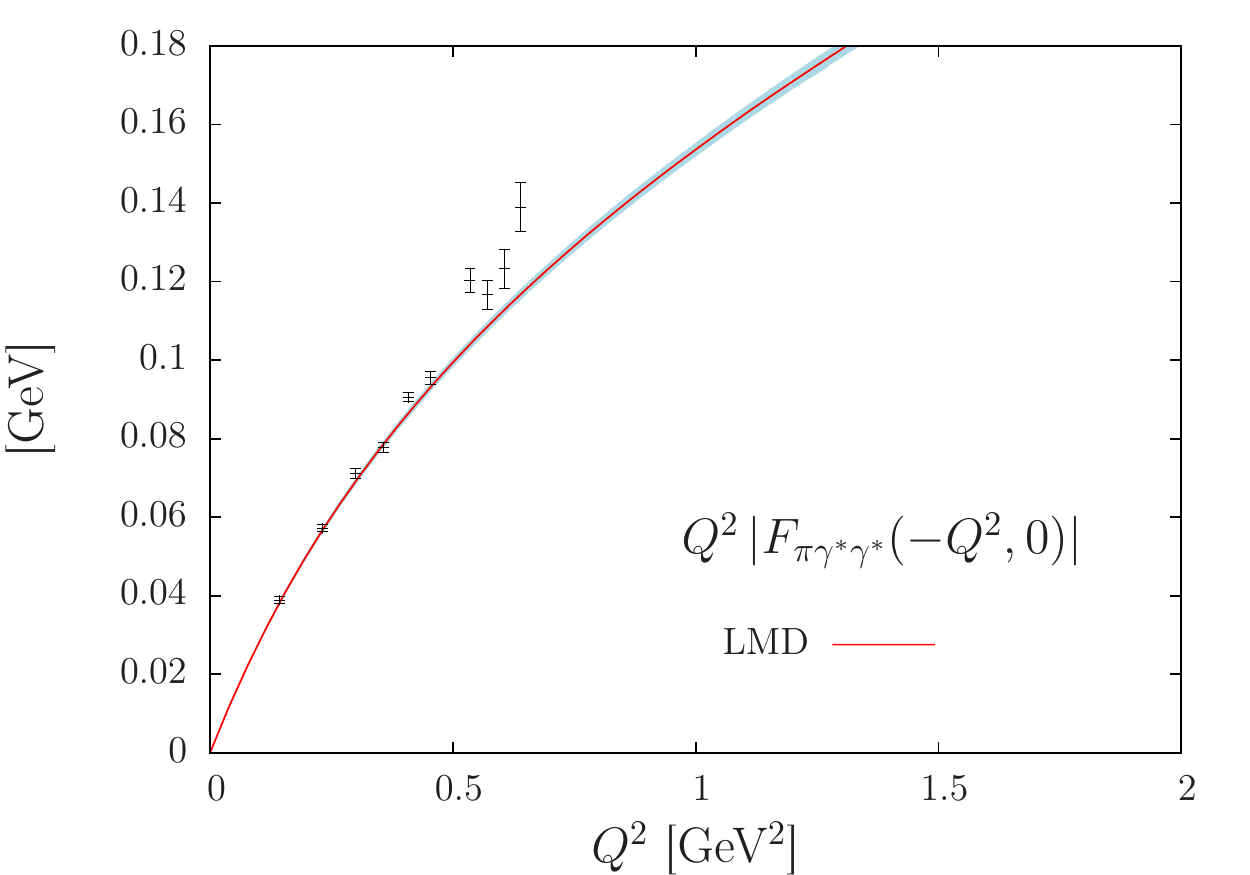}
	\end{minipage}
	\begin{minipage}[c]{0.32\linewidth}
	\centering 
	\includegraphics*[width=0.95\linewidth]{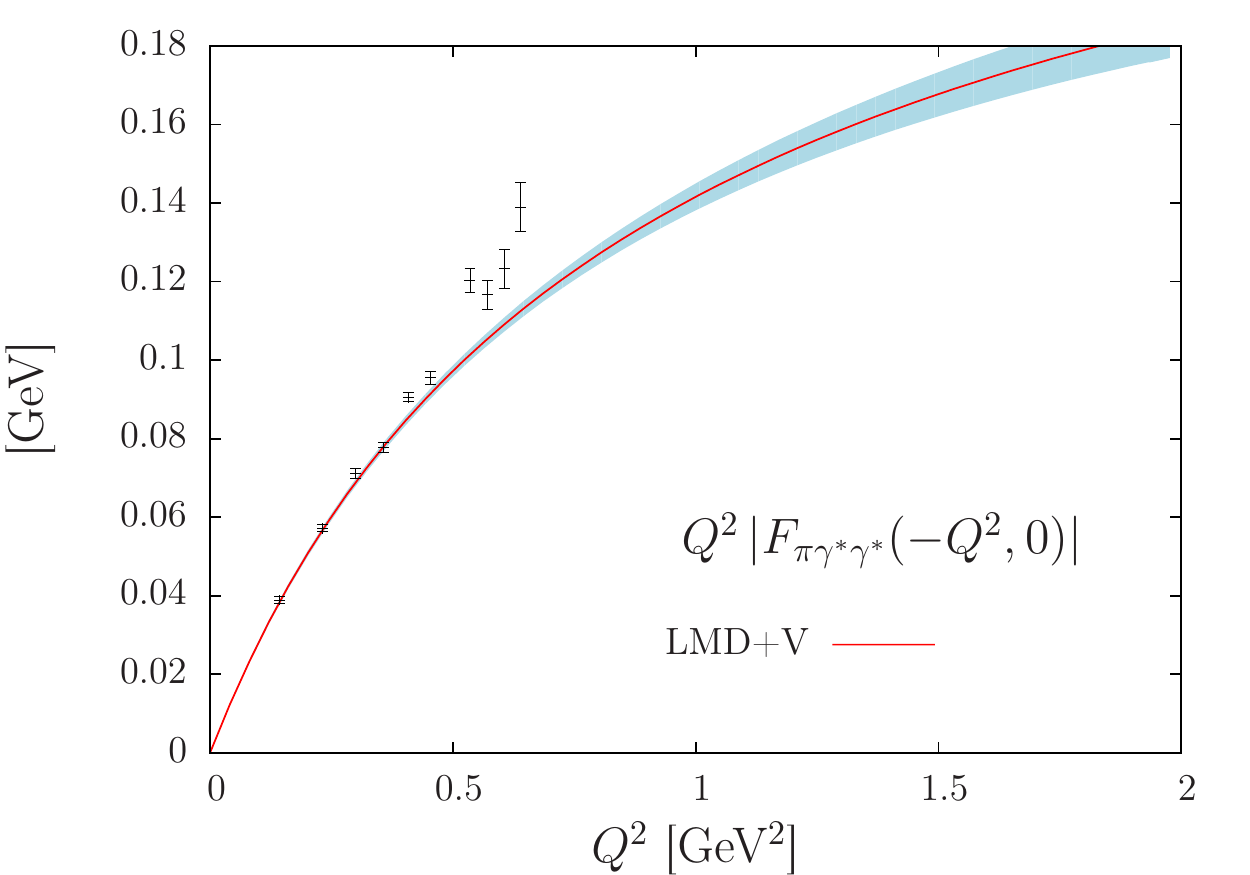}
	\end{minipage}
		
	\caption{Comparison of the VMD, LMD and LMD+V fits for the lattice ensemble O7. The red line corresponds to the results from our global fit. The VMD model falls-off as $\FF^{\rm VMD}(-Q^2, -Q^2) \sim 1/Q^4$ in the double virtual case and fails to describe the numerical data. Note that points at different $Q^2$ are correlated.}	
	\label{fig:fit_ff}
\end{figure}

%
%
We first compare our results with the VMD model, parametrized~by
\begin{equation}
\FF^{\VMD}(q_1^2, q_2^2) = \frac{ \alpha
  M_V^4}{(M_V^2-q_1^2)(M_V^2-q_2^2)} \,.
\label{eq:VMD_model}
\end{equation}
Using $\alpha = 1/(4\pi^2F_{\pi}) = 0.274~\GeV^{-1}$, it reproduces the anomaly constraint in the chiral limit. This model is also compatible with the Brodsky-Lepage behavior (\ref{eq:BL}) in the single-virtual case but falls off faster than the OPE prediction (\ref{eq:OPE}) in the double-virtual case. To reduce the number of fit parameters, a global fit is performed where all lattice ensembles are fitted simultaneously assuming a linear dependence in both $a/a_{\beta=5.3}$ and $\widetilde{y} = m^2_{\pi} / 8 \pi^2 F^2_{\pi}$ for each parameter of the model. We obtain at the physical point 
\begin{equation}
\alpha^{\VMD} = 0.243(18)~\GeV^{-1} \,, \quad M_V^{\VMD} = 0.944(34)~\GeV \,.
\label{eq:resVMD}
\end{equation}
As can be seen in Fig.~\ref{fig:fit_ff}, the VMD model leads to a poor description of our data ($\chi^2/\dof = 2.9$, uncorrelated fit), especially in the double virtual case and at large Euclidean momenta. 
%
%
The second model, the LMD model~\cite{Moussallam:1994xp}, can be parametrized as
\begin{equation}
\FF^{\LMD}(q_1^2, q_2^2) = \frac{ \alpha M_V^4+ \beta(q_1^2
  + q_2^2)}{(M_V^2-q_1^2)(M_V^2-q_2^2)} \,. 
\label{eq:LMD_model}
\end{equation}
Again, this model reproduces the anomaly constraint and is now compatible with the OPE asymptotic behaviour where $\beta = -F_{\pi}/3$ is the theoretical preferred estimate (see Eq.~\ref{eq:OPE}). However, this model does not reproduce the Brodsky-Lepage behavior for the single-virtual form factor given in Eq.~(\ref{eq:BL}). Using $\alpha$, $\beta$ and $M_V$ as free parameters, we now obtain 
\begin{equation}
\alpha^{\LMD} = 0.275(18)(3)~\GeV^{-1} \,, \quad \beta = -0.028(4)(1)~\GeV \,, \quad M_V^{\LMD} = 0.705(24)(21)~\GeV \,,
\label{eq:resLMD_final}
\end{equation}
with $\chi^2 / \dof = 1.3$ (uncorrelated fit) (Fig.~\ref{fig:fit_ff}). The first error is statistical and the second error include systematics as discussed in \cite{Gerardin:2016cqj}. Although this model fails to reproduce the Brodsky-Lepage behavior, it gives a good description of our data  in the considered kinematical range. The anomaly is recovered with a statistical error of 7\% and $\beta$ is compatible with the OPE asymptotic result given in Eq.~(\ref{eq:OPE}). 
%
%
%
Finally, the LMD+V model, proposed in Ref.~\cite{Knecht:2001xc}, includes a second vector resonance and can be parametrized by 
\begin{equation}
\FF^{\LMDV}(q_1^2, q_2^2) = \frac{\widetilde{h}_0\, q_1^2
  q_2^2 (q_1^2 + q_2^2) + \widetilde{h}_1 (q_1^2+q_2^2)^2  +
  \widetilde{h}_2\, q_1^2 q_2^2 + \widetilde{h}_5\, M_{V_1}^2
  M_{V_2}^2\,  (q_1^2+q_2^2) +
  \alpha\, M_{V_1}^4
  M_{V_2}^4}{(M_{V_1}^2-q_1^2)(M_{V_2}^2-q_1^2)
  (M_{V_1}^2-q_2^2)(M_{V_2}^2-q_2^2)} \,. 
\label{eq:LMDV_model}
\end{equation}
One main advantage of this model is that it fulfils all the theoretical constraints discussed in Sec.~\ref{sec:intro} if one sets $\widetilde{h}_1=0$ (which is explicitly done in our fits) and $\widetilde{h}_0 = -F_\pi / 3$. In Ref.~\cite{Knecht:2001xc}, the masses are  set to their physical values $M_{V_1}=m^{\exp}_{\rho} = 0.775~\GeV$ and $M_{V_2}=m^{\exp}_{{\rho}^{\prime}} = 1.465~\GeV$. The parameter ${\widetilde h}_2 = 0.327~\GeV^3$ can be fixed by comparing with the subleading term in the OPE in Eq.~(\ref{eq:OPE}) (Ref.~\cite{MV_04,Novikov:1983jt}) and the parameter ${\widetilde h}_5 = -0.166(6)~\GeV$ has been determined in Ref.~\cite{Knecht:2001xc} by a fit to the CLEO data~\cite{exp} for the single-virtual form factor. 
\begin{figure}[t]

	\vspace{-0.5cm}

	\begin{minipage}[c]{0.49\linewidth}
	\centering 
	\includegraphics*[width=0.85\linewidth]{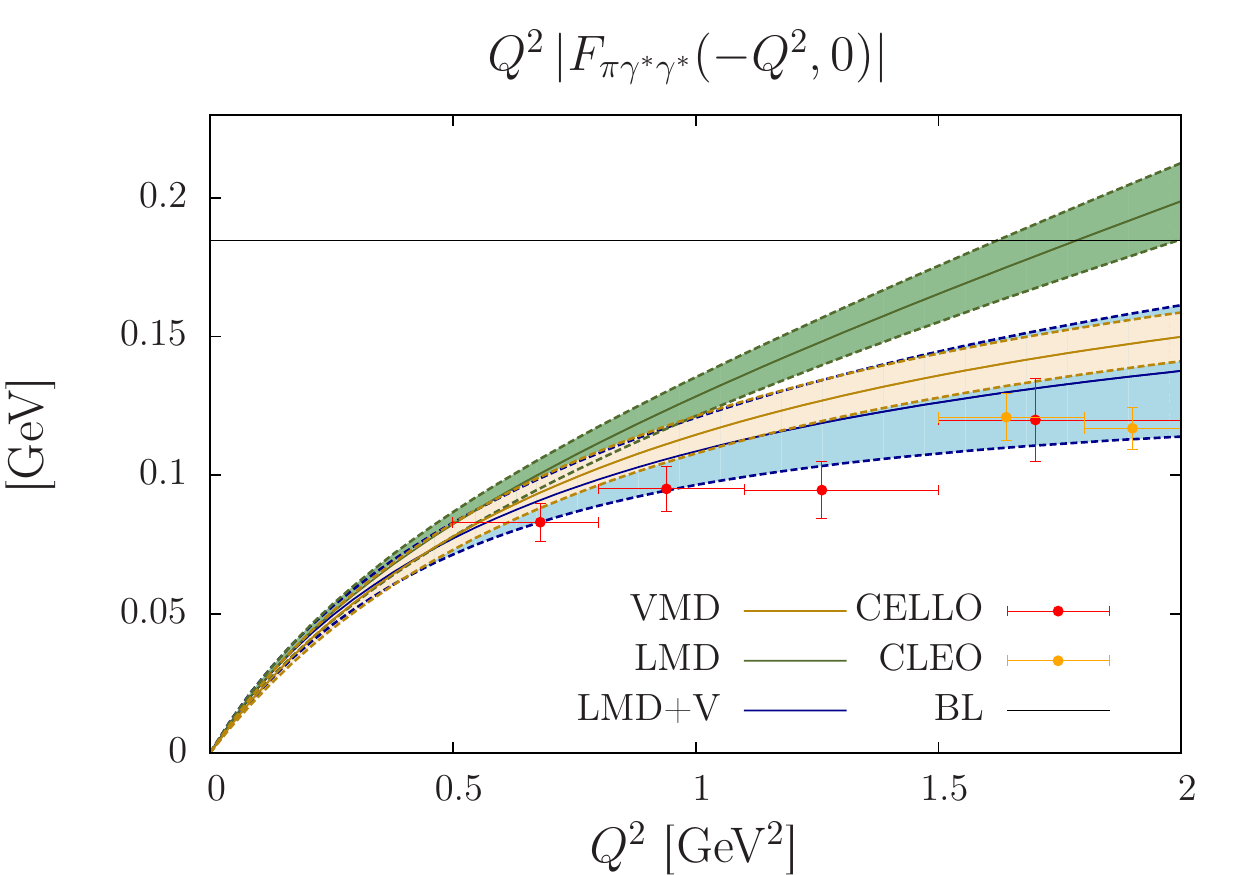}
	\end{minipage}
	\begin{minipage}[c]{0.49\linewidth}
	\centering 
	\includegraphics*[width=0.85\linewidth]{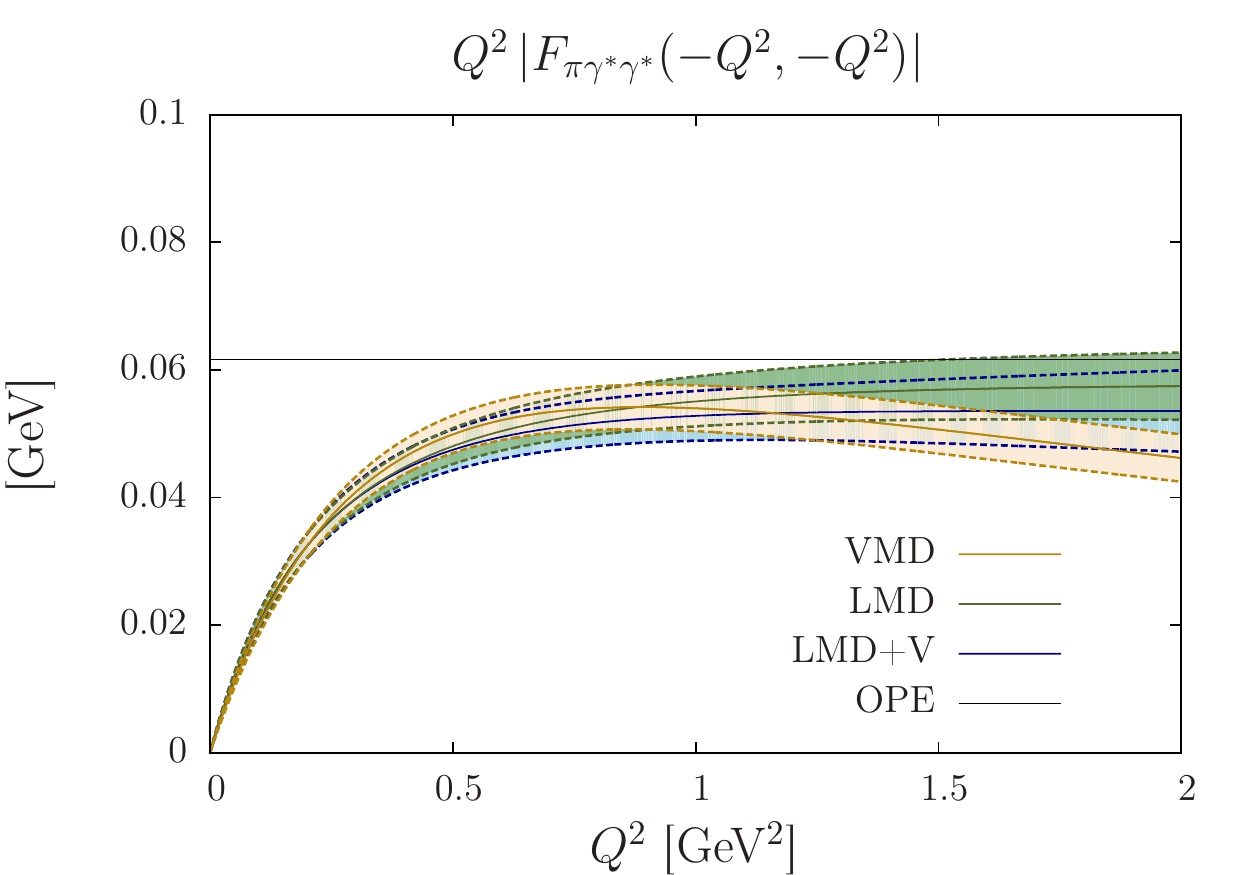}
	\end{minipage}
	
	\caption{Lattice extrapolations for the VMD, LMD and LMD+V models. (left) Single-virtual form factor. 
	 (right) Double-virtual form factor at $Q_1^2=Q_2^2$.}

 	\label{fig:FF_cmp}
\end{figure}
To get stable fits, we enforce the constraint $M_{V_1} = m_{\rho}^{\exp}$ at the physical point but still allowing for chiral corrections. For $M_{V_2}$, inspired by quark models, we assume a constant shift in the spectrum and set $M_{V_2}(\widetilde{y}) = m^{\exp}_{\rho^{\prime}} + M_{V_1}(\widetilde{y}) - m^{\exp}_{\rho}$. Finally, we impose the theoretical constraint $\widetilde{h}_0 = -F_{\pi}/3$ in the continuum and chiral limit but, again, still allowing for chiral and lattice artefacts corrections. Using these assumptions, we obtain
\begin{equation}
\alpha^{\LMDV} = 0.273(24)(7)~\GeV^{-1} \,, \ \widetilde{h}_2 = 0.345(167)(83)~\GeV^3 \,, \ \widetilde{h}_5 = -0.195(70)(34)~\GeV \,,
\label{eq:resLMDV1_final}
\end{equation}
with $\chi^2 / \dof = 1.4$ (uncorrelated fit). This model also gives a good description of our data as can be seen in Fig.~\ref{fig:fit_ff} and turns out to be close to the LMD model in the kinematical range considered here. The systematic error has been estimated by varying our assumptions on $M_{V_1}$ and $M_{V_2}$. Again, the anomaly constraint is recovered within statistical error bars and the values of $\widetilde{h}_2$ and $\widetilde{h}_5$ are in good agreement with phenomenology. 

The form factor extrapolated to the physical point for each model is shown in Fig.~\ref{fig:FF_cmp}. In the single-virtual case, the VMD and LMD+V models are in good agreement with the experimental data whereas the LMD model starts to deviate at $Q^2=1~\GeV^2$. In the double-virtual case, the LMD and LMD+V models are similar and already close to their asymptotic behavior at $Q^2 \sim 1.5~\GeV^2$ where we have lattice data. Finally, using the formalism developed in Ref.~\cite{Jegerlehner:2009ry} and our result for the form factor, we estimate the pion-pole contribution $\amu$ to hadronic light-by-light scattering in the muon $g-2$. Our preferred estimate for $\amu$ is obtained with the fitted LMD+V model \cite{Gerardin:2016cqj},  
\begin{equation} \label{amupi0LMDV_lattice}
a_{\mu; \LMDV}^{\mathrm{HLbL}; \pi^0} = (65.0 \pm 8.3) \times 10^{-11} \,.  
\end{equation} 
For comparison, most model calculations yield results in the range $\amu = (50-80) \times 10^{-11}$ with rather arbitrary, model-dependent error estimates, see Refs.~\cite{Jegerlehner:2009ry,Bijnens:2015jqa,KN_02_Nyffeler:2016gnb} and references therein.

\begin{acknowledgments}
\vspace{-0.3cm}
We are grateful for the access to the lattice ensembles, made available to us through CLS.  We acknowledge the use of computing time for the generation of the gauge configurations on the JUGENE and JUQUEEN computers of the Gauss Centre for Supercomputing located at Forschungszentrum J\"ulich, Germany. All correlation functions were computed on the dedicated QCD platforms ``Wilson'' at the Institute for Nuclear Physics, University of Mainz, and ``Clover'' at the Helmholtz-Institut Mainz. This work is partly supported by the DFG through CRC 1044.
\end{acknowledgments}

\end{document}